\documentclass[aps,prd,reprint,showpacs,superscriptaddress,twocolumn]{revtex4-2}
\usepackage{amsfonts}
\usepackage{amsmath}
\usepackage{amssymb}
\usepackage{dsfont}
\usepackage{hyperref}
\usepackage{graphicx}
\usepackage{float}
\usepackage[usenames,dvipsnames]{xcolor}
\usepackage[normalem]{ulem}
\usepackage{times}
\usepackage{braket}
\usepackage{booktabs}
\usepackage{multirow}
\usepackage{subcaption}
\usepackage{tikz}
\usepackage[font=normal, justification=Justified]{caption}
\hypersetup{
  colorlinks=true,linkcolor=blue,citecolor=blue,
  filecolor=blue,urlcolor=blue,breaklinks=true
}

\begin{document}

\title{Effects of rotation on the thermodynamic properties of a quantum dot}

\author{Lu\'{i}s Fernando C. Pereira}
\email{luisfernandofisica@hotmail.com}
\affiliation{Departamento de F\'{\i}sica, Universidade Federal do Maranh\~{a}o, 65085-580 S\~{a}o Lu\'{\i}s, Maranh\~{a}o, Brazil}

\author{Edilberto O. Silva}
\email{edilberto.silva@ufma.br}
\affiliation{Departamento de F\'{\i}sica, Universidade Federal do Maranh\~{a}o, 65085-580 S\~{a}o Lu\'{\i}s, Maranh\~{a}o, Brazil}

\date{\today}

\begin{abstract}
In this work, we investigate the effects of rotation on the physical properties of a quantum dot described by a radial potential and subjected to a rotating reference frame. The interplay between rotation and confinement is analyzed by solving the Schr\"{o}dinger equation for the system, yielding energy levels and wavefunctions as functions of angular velocity. We compute key thermodynamic properties, including the density of states, magnetization, entropy, and heat capacity, in the absence of an external magnetic field. Our results demonstrate that rotation induces significant modifications to the energy spectrum, removing degeneracies and generating oscillatory behaviors in magnetization akin to de Haas-van Alphen and Aharonov-Bohm-type oscillations. Furthermore, we observe an effect analogous to the magnetocaloric effect, where an increase in angular velocity leads to a decrease in temperature during adiabatic processes. These results reveal the potential of rotational effects to influence quantum systems and provide insights for future studies in mesoscopic physics.
\end{abstract}

\maketitle

\section{Introduction}
\label{intro}

Quantum dots are nanoscale structures where electrons are confined in all three spatial dimensions, creating discrete energy levels reminiscent of atomic systems \cite{tartakovskii2012quantum,chakraborty1999quantum,unlu2022progress}. These systems provide a versatile platform for investigating quantum mechanical phenomena, such as electron-electron interactions and spin dynamics, with numerous applications in quantum computing, optoelectronics, and nanophotonics \cite{PRR.2022.4.033043,PRB.2021.104.235302,JMMM.2018.449.510,CJP.2024.92.824,ACS.2024.128.2506}. The tunable properties of quantum dots, achieved via external fields or changes in confinement potentials, enable detailed studies of many-body physics and strongly correlated systems.

Building upon this foundational knowledge, recent research has explored rotating effects on quantum systems \cite{Anandan2004,PLA.2015.379.11,RSOS.2017.4.170541,PRB.2011.84.104410,PBCM.2006.385.1381,FP.1973.3.493,FP.1974.4.75,FP.1980.10.151,PLA.1993.181.421,PLA.1995.197.444,PRB.1996.54.1877,PA.1996.233.503,EL.2001.54.502,AofP.2014.346.51,PLA.2018.36.2499,universe.2021.7.457,FBS.2022.63.58,AdP.2023.535.202200371,EPJP.2023.138.118,AP.2023.459.169547,CTP.2024.76.105701,QR.2024.6.6040041,universe.2024.10.10100389,CinTP.2024.76.065504}. For instance, Ref. \cite{JP.2007.19.076211} investigated the spectral properties of rotating electrons in quantum dots within the lowest Landau level using exact diagonalization. Their findings revealed transitions between spin-polarized and partially polarized states, the emergence of spin waves, and electron localization at high angular momentum. These results connect quantum dot behavior to quantum Hall liquid phenomena and provide valuable insights into strongly correlated systems and their potential applications in nanotechnology and quantum computing.

Theoretical models have further illuminated the influence of rotation on spin-polarized systems. For example, Ref. \cite{PRR.2020.2.033492} demonstrated that rotation induces spin currents and modifies spin textures in two-dimensional electron gases, offering potential applications in spintronics. Additionally, Ref. \cite{AofP.2019.407.154} employed non-Hermitian approaches to analyze the dynamics of rotating quantum systems, unveiling new behavior regimes in dissipative quantum mechanics.

In this work, we focus on studying the effects of rotation on the thermodynamic properties of quantum dots, with particular attention to the density of states, which serves as a key parameter in our analysis. Using the approach presented in Ref. \cite{QR.2024.6.664}, we calculate the density of states for a quantum dot under rotation, which, in turn, allows us to derive integral equations for thermodynamic properties such as entropy, magnetization, and heat capacity. Among the effects of rotation,  a centrifugal potential plays a significant role. For parabolic confinement, as shown by Bencheikh \textit{et al.} \cite{PRA.2014.89.063620}, this potential imposes restrictions on angular velocity. Here, we generalize this analysis to a model incorporating parabolic confinement and a repulsive potential, revealing new constraints and behaviors.

The structure of our article is as follows. In Section \ref{sec:model}, we describe the theoretical model used to analyze the effects of rotation on the quantum dot, detailing the potential and confinement terms. Section \ref{Sec:properties} presents the calculation of the density of states and other thermodynamic properties as functions of magnetic field, temperature, and angular velocity. Numerical results and their implications are discussed in Section \ref{Sec:Resultados}, focusing on how rotation modifies physical properties and induces effects analogous to magnetic phenomena. Finally, Section \ref{Conc} summarizes the results and outlines potential directions for future research.

\section{Description of the model}
\label{sec:model}

In this section, we describe the model of an electron with effective mass $m^{*}$ and charge $e$ confined by a radial potential  $V\left(r\right)$ — the system in a rotating frame in a uniform magnetic field. 

We assume that $\boldsymbol{\omega}=\omega\mathbf{\hat{z}}$ and $\mathbf{B}=B\mathbf{\hat{z}}$, where
$\omega$ is the angular velocity of the rotating frame and $B$ represents the magnetic field strength, respectively. In this case,
the motion of the particle governed by the Schr\"{o}dinger equation is given by
\begin{equation}
\left[\frac{1}{2m^{*}}\left(\mathbf{p}-e\mathbf{A}_{1}-m^{*}\mathbf{A}_{2}\right)^{2} +V_{c}\left(r\right) +V(r) \right]
\psi = E\psi, 
\label{Eq:eq.Sch.2D}
\end{equation}
where $\mathbf{A}_{1}=\left(Br/2\right)\hat{\boldsymbol{\varphi}}$ is the symmetric gauge vector   potential, $\mathbf{A}_{2}=\omega r\hat{\boldsymbol{\varphi}}$ is the gauge for the rotating frame, and 
\begin{equation}
V_{c}\left(r\right)=-\frac{1}{2}m^{*}\omega^{2}r^{2},
\label{Eq:centrifugo}
\end{equation}
is a contribution to the centrifugal potential energy in the rotating frame. 
Additionally, the radial potential is given by \cite{SST.1996.11.1635} 
\begin{equation}
	V\left( r\right)=\frac{a_{1}}{r^{2}}+a_{2}r^{2}-V_{0},
	\label{PotRad}
\end{equation}
where the first term is a repulsive potential, the second term corresponds to a parabolic confining potential, which restricts the wave functions to a finite region, and $V_{0}=2\sqrt{a_{1}a_{2}}$.
The radial potential has a minimum at $r_{0}=\left(a_{1}/a_{2}\right)^{1/4}$. Furthermore, it can be shown that $\omega_{0}=\sqrt{8a_{2}/m^{*}}$, which defines the strength of the transverse confinement. It is known that any variation in parameters $a_{1}$ and $a_{2}$ implies changes in the spatial distribution of the electron. The width of the system at a given Fermi energy $E_{F}$ can be estimated by
\begin{equation}
\Delta R=\sqrt{\frac{8E_{F}}{m^{*}\omega_{0}^{2}}}.
\label{Eq:raio}
\end{equation}

The radial potential serves as a model for the theoretical definition of a ring of average radius $r_{0}$ with finite width. However, it can also describe other systems, such as a quantum dot with $a_{1}=0$ and an anti-dot with  $a_{2}=0$. The radial potential creates an environment where the electron's quantum properties are highlighted, giving rise to phenomena such as confinement and discrete energy level structures. 

The energy eigenvalues and wavefunctions of Eq. (\ref{Eq:eq.Sch.2D}) are given, respectively, by
\begin{equation}
	E_{n,m}=\left(n+\frac{1}{2}+\frac{L}{2}\right) \hbar\omega_{1} -\frac{m}{2}\hbar \omega_{2}-V_{0},  
	\label{Eq:Enm}
\end{equation}
and
\begin{align}
	\psi_{n,m} (r,\varphi ) =& \frac{1}{\lambda }\sqrt{\frac{\Gamma\left[n+L+1\right]}{2\pi n!\left(\Gamma\left[L+1\right]\right)^{2}}}
	e^{im\varphi }e^{ -\frac{r^{2}}{4\lambda ^{2}}}\left( \frac{r^{2}}{2\lambda ^{2}}\right) ^{\frac{L}{2}}
	\notag \\
	& \times{\mathrm{M}\left( -n,\,1+L ,\frac{r^{2}}{2\lambda ^{2}}\right)},
	\label{Eq:funcaodeonda}\end{align}
where $n=0,1,2,\ldots$ is the radial quantum number, which denotes a subband, $m=0,\pm 1,\pm2,\ldots$ is the magnetic quantum number, $L=\sqrt{m^{2}+2a_{1}m^{*}/\hbar^{2}}$
is the effective angular momentum, and $\lambda=\sqrt{\hbar/m^{*} \omega_{1}}$ is the effective magnetic length, and we have defined $\omega_{1} = \sqrt{\omega_{2}^{2}
+\omega_{0}^{2}-4\omega^{2}}$, where $\omega_{2}=\omega_{c}+2\omega$ and $\omega_{c}=eB/m^{*}$ is the cyclotron frequency. Also, $\mathrm{M}\left(a,c,x\right)$ is the confluent hypergeometric function of the first kind. 

Among the features of the model, we can highlight that for a given magnetic field  and angular velocity, the minimum energy of all subbands is in the same state  $m=m_{0}$ given by
\begin{equation}
m_{0}=\sqrt{\frac{2a_{1}m^{*}}{\hbar^{2}}\frac{\omega_{2}^{2}}{\omega_{0}^{2}-4\omega^{2}}},
\label{Eq:m0}
\end{equation}
as long as $0 \leq \omega < \omega_{0}/2$. 

The result expressed by Eq. (\ref{Eq:m0}) shows an interesting aspect of the model: for angular velocity positive values outside the range defined above, the confinement imposed on the system by the parabolic potential is eliminated. This result is a consequence of the presence of the centrifugal potential.
In particular, for $\omega= \omega_{0}/2$, resonance occurs between the angular velocity and the
natural frequency of the confinement. In this case, the system assumes the characteristics of a usual anti-dot centered at $r=r_{0}$. Furthermore, if we eliminate the repulsive potential by making $a_{1}=0$,
the energy levels exhibit infinite degeneracy; in this configuration, we obtain the rotational analog of the degenerate Landau levels. 

\section{Physical properties}

\label{Sec:properties}

In this work, we focus on the case of a quantum dot. The results of Section \ref{sec:model} show that we should consider only the angular velocity values in the interval $0 \leq \omega < \omega_{0}/2$. The number of states at energy $E$ in a subband $n$ is given by
\begin{equation}
	N_{n}=\frac{\Delta N}{\hbar \omega_{1} }\left(E-E_{n,0}\right)\;\Theta\left(E-E_{n,0}\right),  
	\label{Eq:Nn}
\end{equation}%
where $\Theta\left(E-E_{n,0}\right)$ is the Heaviside function, and $\Delta N$ corresponds to the number of states in energy interval $E_{n,0}\leq E < E_{n+1,0}$, given by
\begin{equation}
	\Delta N=\frac{4\omega _{1}^{2}}{\omega _{0}^{2}-4\omega ^{2}}.  
	\label{Eq:dNn}
\end{equation}

Using Eq. (\ref{Eq:Nn}), we compute the density of states of a subband:
\begin{equation}
D_{n}\left(E\right)=\frac{\Delta N}{\hbar \omega_{1}} \;\Theta\left(E-E_{n,0}\right),
\label{Eq:DOS_QD}
\end{equation}
The total density of states takes a ladder-like function. Each subband contributes a step of
magnitude $\Delta N/\hbar \omega_{1}$. 
Using the density of states, we can analyze various physical properties as a function of the magnetic field, angular velocity, and temperature $T$ at a sample with a fixed particle number $N$.

To analyze the properties of a quantum dot, we must know the chemical potential $\mu$ as a function of temperature, magnetic field, and angular velocity. It is computed, self-consistently, from the expression
\begin{equation}
	N=\displaystyle \sum_{n=0}^{\infty}\int_{0}^{\infty} f\left(E,\mu \right) D_{n}\left(E\right)dE.
	\label{Eq:N}
\end{equation}
where  $f\left(E,\mu\right)$ is the Fermi-Dirac distribution
function, given by
\begin{equation}
	f\left(E,\mu\right)=\frac{1}{1+e^{\frac{E-\mu}{k_{B}T}}},
	\label{Eq:distribuicao}
\end{equation}
with $k_{B}$ the Boltzmann constant. Given the simple form in which the density of states is presented (see Eq. (\ref{Eq:DOS_QD})), we can rewrite Eq. (\ref{Eq:N}) as
\begin{equation}
	N=k_{B}T\displaystyle \sum_{n=0}^{\infty}D_{n}\left( E\right) \ln \left( 1+e^{\frac{\mu -E}{k_{B}T}}\right).
	\label{Eq:N2}
\end{equation}

At $T=0$, we can define  the Fermi energy as 
\begin{equation}
E_{F}=\left(n_{max}+\delta+\frac{1}{2}\right)\hbar\omega_{1},  
\label{Eq:Fermi}
\end{equation} 
where $n_{max}$ corresponds to the highest occupied subband at the Fermi energy, and $0 \leq \delta < 1$. The $n_{max}$ and $\delta$ parameters  can be obtained from Eq. (\ref{Eq:Nn}), and are given by 
\begin{equation}
n_{max}\equiv [x],\; x=\sqrt{\frac{1}{4}+2\nu}-\frac{1}{2},
\label{Eq:nmax}
\end{equation}
where $[x]$ denotes the largest integer less than or equal to $x$, $\nu=N/\Delta N$, and  
\begin{equation}
\delta =\frac{\nu}{n_{max }+1}-\frac{n_{max}}{2}. 
\label{Eq:delta}
\end{equation}

Knowing the chemical potential, we can obtain the free energy $F$,
\begin{equation}
F=N\mu -k_{B}T\displaystyle \sum_{n=0}^{\infty}\int_{0}^{\infty}\ln \left(
1+e^{\frac{\mu -E}{k_{B}T}}\right) D_{n}\left( E\right) dE,
\label{Eq:EnergiaLivre}
\end{equation}
of which we can compute the magnetization $M=-\left( \partial F/\partial B \right)_{T,N}$, 
\begin{align}
M=&-\frac{e\hbar }{m^{\ast }}\frac{\omega _{2}}{\omega _{1}}\frac{1}{\hbar\omega_{1}}\Bigg[ \displaystyle \sum_{n=0}^{\infty}E_{n,0}\int_{0}^{\infty }f\left(E,\mu\right)D_{n}\left( E\right)dE \notag \\
&-k_{B}T\displaystyle \sum_{n=0}^{\infty}\int_{0}^{\infty }\ln \left( 1+e^{\frac{\mu -E}{k_{B}T}}\right) D_{n}\left(
E\right) dE\Bigg],
\label{Eq:magnetization}
\end{align}
and entropy $S=-\left( \partial F/\partial T \right)_{B,N}$,
\begin{equation}
S=k_{B}\displaystyle \sum_{n=0}^{\infty} \int_{0}^{\infty}
\left[ \ln \left( 1+e^{\frac{\mu -E}{k_{B}T}}\right) +\frac{\frac{E-\mu }{k_{B}T}}{e^{\frac{E-\mu}{k_{B}T}}+1}\right]D_{n}\left( E\right) dE.
\label{Eq:entropia}
\end{equation}

The heat capacity is given, in general, as
\begin{equation}
	C_{e}=\displaystyle \sum_{n=0}^{\infty}\int_{0}^{\infty} \frac{df\left(E,\mu \right)}{dT} D_{n}\left(E\right)\left(E-\mu\right)dE.
	\label{Eq:calorespecifico}
\end{equation}
Following the same procedures as in Ref. \cite{SSC.1984.50.537}, we can show that the heat capacity can be written as 
\begin{equation}
	C_{e}=k_{B}\left(L_{2}-\frac{L_{1}^{2}}{L_{0}}\right),
	\label{Eq:calorespecifico2}
\end{equation}
where we define the parameter $L_{r}$ as
\begin{equation}
	L_{r}=\displaystyle \sum_{n=0}^{\infty}\int_{0}^{\infty}\frac{e^{\frac{E-\mu}{k_{B}T}}}{\left(1+e^{\frac{E-\mu}{k_{B}T}}\right)^{2}}\left(\frac{E-\mu}{k_{B}T}\right)^{r}D_{n}\left( E\right)dE.
	\label{Eq:L}
\end{equation}
 
\section{Numerical results}
\label{Sec:Resultados} 

For numerical analysis, we consider a sample made of GaAs with the effective mass of the electron $m^{*}=0.067 m_{e}$, where $m_{e}$ is the electron mass. We also consider that there are $N=1400$ spinless electrons.
The confinement energy parameter used for this purpose is $\hbar \omega_{0}=0.459$ meV. Furthermore, let us consider the case where there is only rotation, i.e., with the magnetic field turned off.
When, in addition to the rotation, there is a magnetic field, we can show that this corresponds to an increase in the effects of the magnetic field on the electronic states. This is evident given the way the frequencies $\omega_{1}$ and $\omega_{2}$ are defined: $\omega_{1} = \sqrt{\omega_{2}^{2}
+\omega_{0}^{2}-4\omega^{2}}$, $\omega_{2}=\omega_{c}+2\omega$. Therefore, rotation evidences the effects of the magnetic field in two ways: directly by increasing the magnetic field and indirectly by decreasing the effects of transverse confinement. 
In this context, a vast amount of literature deals with the magnetic field's effects on a quantum dot's physical properties. So, we will focus on the case in which the magnetic field is off.  With this in mind, we write the angular velocity in terms of transverse confinement as  $\omega=\varpi\omega_{0}$, where $\varpi$ is the relative frequency. This work considers only positive angular velocity values so that $0 \leq \varpi < 1/2$. 

\begin{figure}[!t!]
	\centering	
	\includegraphics[scale=0.48]{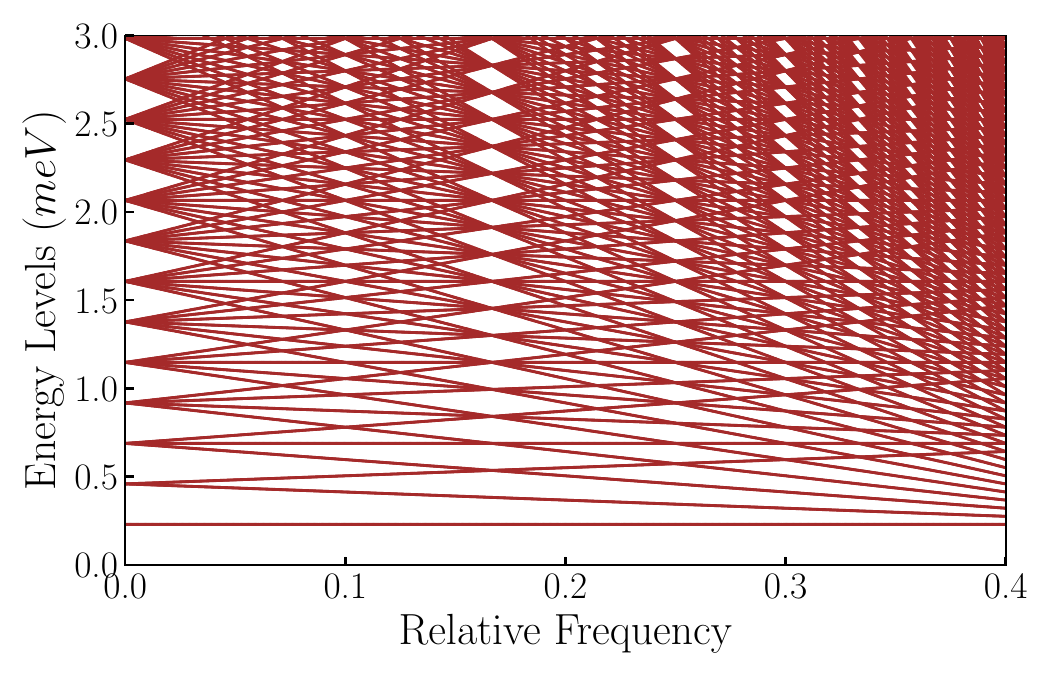}
	\caption{Energy levels of a quantum dot as a function of relative frequency.}
	\label{Fig:Energy}
\end{figure}

Figure \ref{Fig:Energy} exhibits the evolution of energy levels as a function of relative frequency. The states with $m \neq 0$ are linear functions of the angular velocity; the states with $m<0$ and $m>0$ are increasing and decreasing functions, respectively. The states with $m=0$ do not depend on the angular velocity. As is known in the literature, the states of a quantum dot are degenerate, except the state with $n=m=0$ (we are not considering spin-degeneracy). Rotation removes degeneracy. But as the angular velocity
increases, accidental degeneracies may occur, again leading to
for example, the enhanced bunching of single-particle levels at $\varpi=1/6, 1/4,$ etc.

\begin{figure}[!t!]
\centering	
\includegraphics[scale=0.45]{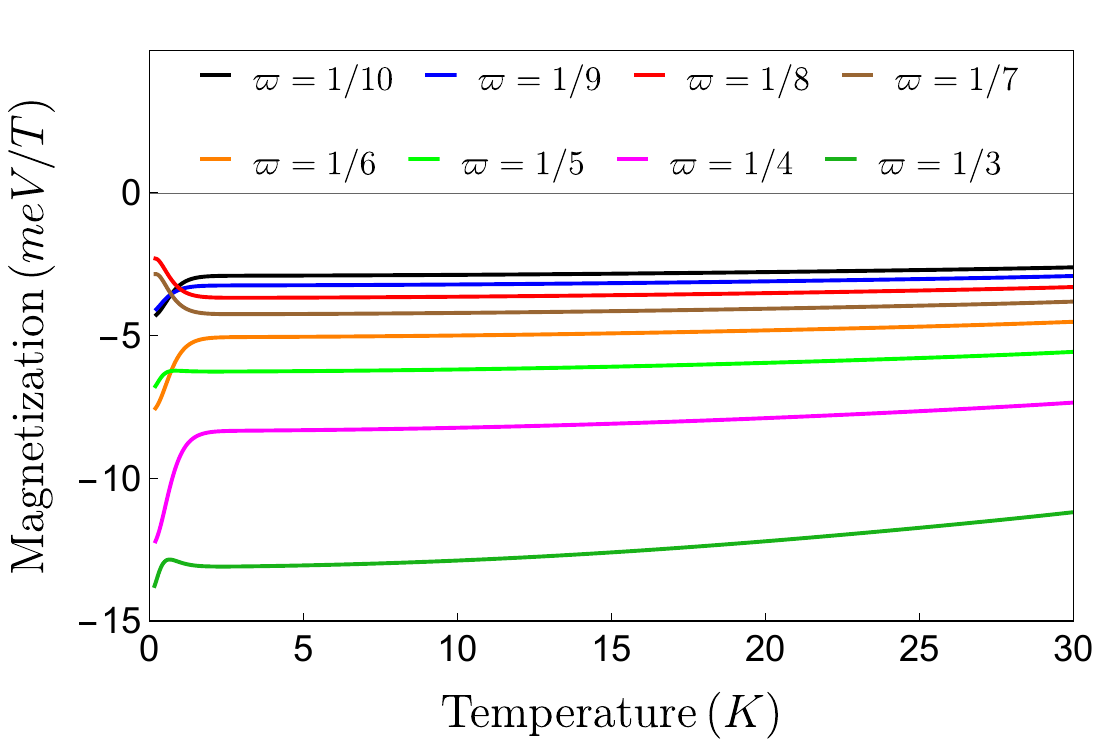}
\caption{Magnetization of a quantum dot as a temperature function for different relative frequency values.}
\label{Fig:Magnetization}
\end{figure}

Figure \ref{Fig:Magnetization} shows the magnetization as a function of temperature for different values of relative frequency $\varpi$. Interestingly, even without magnetic fields, the quantum dot still exhibits magnetization due to angular velocity. This is a manifestation of the Barnett effect, namely, uniform rotation of a body causes a magnetization \cite{PR.1915.6.239,AJofP.1948.16.140,book.landau.vol4,APL.2009.95.122504,Nature.2024.628.540}.
On the other hand, a classical result shows that a disk of radius $R$ with total charge $Q$ distributed uniformly over its surface and rotates with angular velocity $\omega$, produces a magnetic field. For a point $z$ on an axis perpendicular to the plane of the disk and passing through its center, this magnetic field is given by \cite{griffiths2014introduction} 
\begin{equation}
B=\frac{\mu_{0}\sigma\omega}{2}\left(\frac{R^{2}+2z^{2}}{\sqrt{R^{2}+z^{2}}}-2z\right),
\label{Eq:Binduzido}
\end{equation}
where $\mu_{0}$ is the magnetic constant and $\sigma$ is the surface charge density.
This result, although classical, can give us an estimate of the magnetic field induced by the rotation of a quantum dot. Generally, the electron concentration in a quantum dot described by a parabolic potential increases from the edge to the center, so the charge density is not uniform. However, for simplicity, we assume that the charge density is uniform. In our model, $Q=Ne$, and the radius of the quantum dot is calculated from Eq. (\ref{Eq:raio}). Note that by setting the number of electrons fixed, the Fermi energy and the quantum dot radius must be a function of the system variables. Using the above information and the relative frequency values used in Fig. \ref{Fig:Magnetization}, we can show that the induced magnetic field at the center of the quantum dot is of the order of $10^{-5}$ to $10^{-4}$ T. 
Another essential point in Fig. \ref{Fig:Magnetization} is the profile of the curves in the temperature range below $3.0$ K, which indicates the presence of magnetization oscillations as a function of angular velocity. As we can infer from Fig. \ref{Fig:Energy}, the rotation induces the depopulation of subbands and the crossing of states.  It is a well-known fact in the literature that a magnetic field also causes these behaviors in energy levels and, consequently, induces the appearance of dHvA-type and AB-type oscillations in some physical properties, such as magnetization.
From this similarity, we can say that rotation also induces dHvA-type and AB-type oscillations. The oscillations disappear for temperatures above $3.0$ K.

\begin{figure}[!t!]
	\centering	
	\includegraphics[scale=0.45]{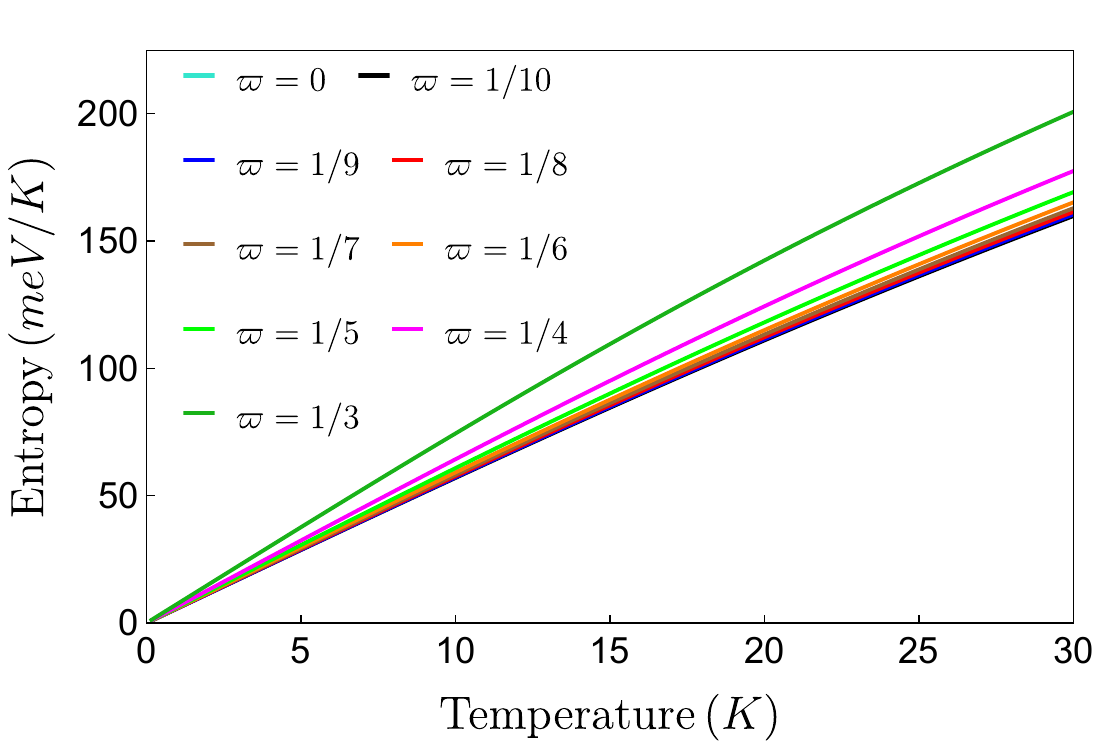}
	\caption{Entropy of a quantum dot as a temperature function for different relative frequency values.}
	\label{Fig:Entropy}
\end{figure}

\begin{figure}[!t!]
	\centering	
	\includegraphics[scale=0.48]{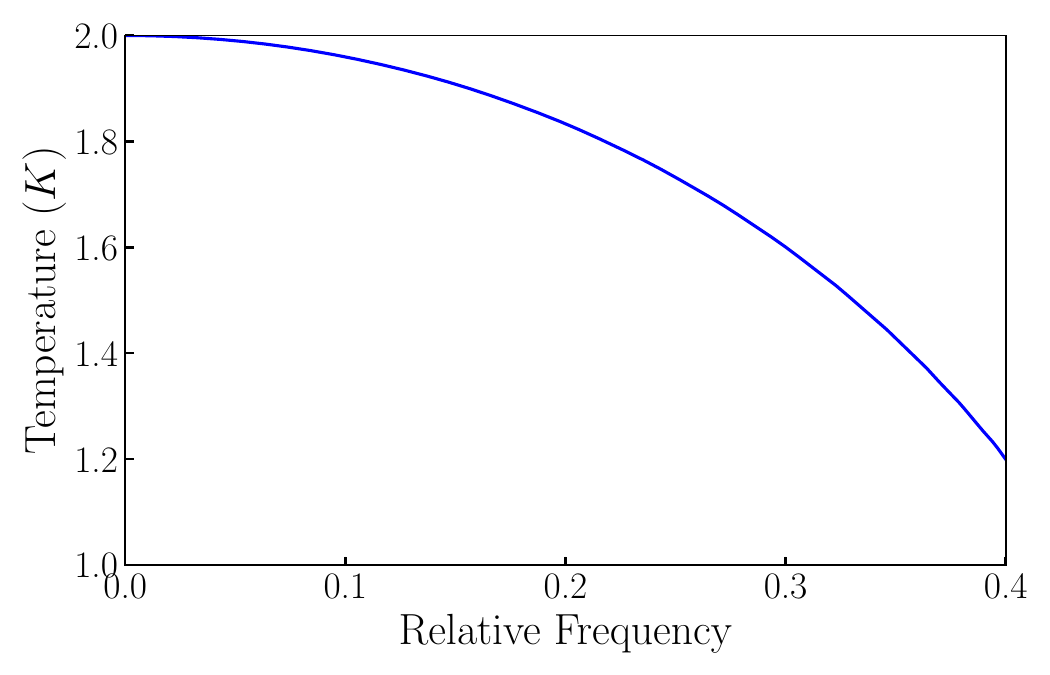}
	\caption{In an adiabatic process, the temperature of a quantum dot is a function of relative frequency.}
	\label{Fig:Temperatura}
\end{figure}

In an adiabatic process, there is no energy transfer either into or out of a system. Since there is no change in energy, the system's entropy remains constant. Since the system variables are temperature and angular velocity, a relationship between them must exist for entropy to remain constant. Figure \ref{Fig:Entropy} shows that this relationship is such that an increase in angular velocity implies a decrease in temperature. In other words, the increase in disorder caused by the increase in angular velocity is compensated by the decrease in disorder caused by the decrease in temperature.  Figure \ref{Fig:Temperatura} shows the temperature behavior in an adiabatic process when the quantum dot has a nonzero angular velocity. Therefore, rotation induces an effect analogous to the magnetocaloric effect. 
 
\begin{figure}[!t!]
	\centering	
	\includegraphics[scale=0.45]{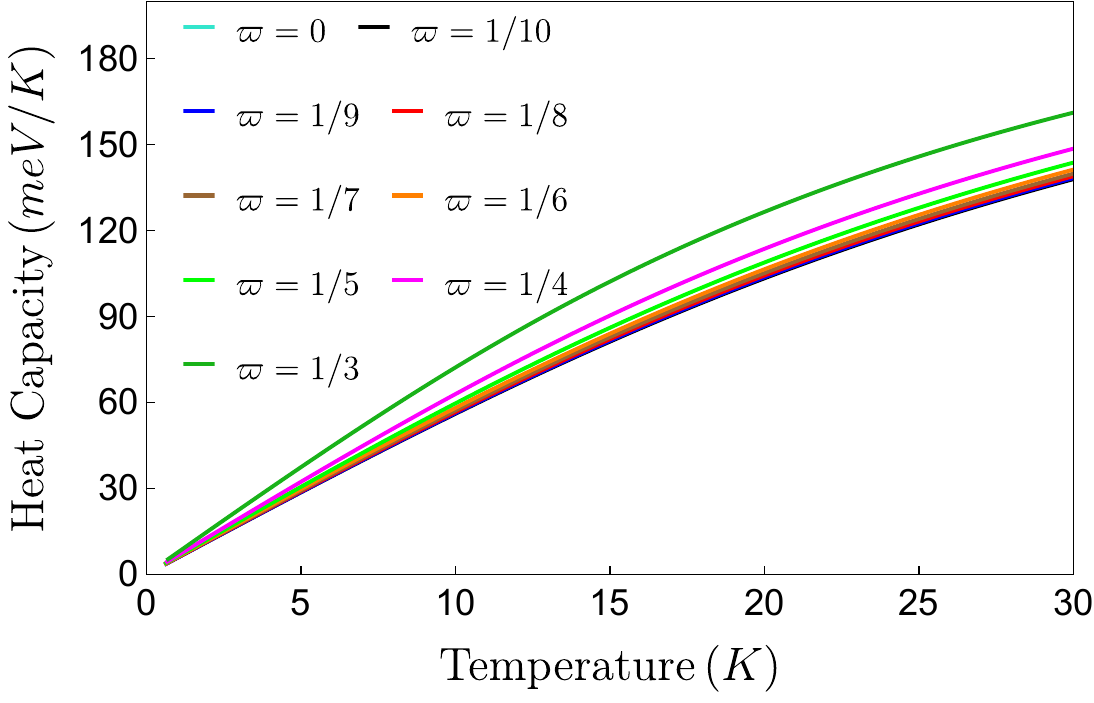}
	\caption{Heat capacity of a quantum dot as a temperature function for different relative frequency values.}
	\label{Fig:HeatCapacity}
\end{figure}  
 
Finally, knowing how the system exchanges energy with its surroundings is important. Figure \ref{Fig:HeatCapacity} shows the behavior of a quantum dot's heat capacity as a temperature function for various relative frequency values. At low temperatures, the effect of rotation is small. As the temperature increases, a more significant rotation contribution to the sample's heat capacity is observed.

\section{Conclusions}
\label{Conc}
We have analyzed the effects of rotation on the physical properties of a quantum dot modeled by a radial confinement potential. By solving the Schr\"{o}dinger equation in a rotating reference frame, we derived the energy spectrum and wavefunctions of the system, highlighting the role of angular velocity in lifting degeneracies and modifying the spatial distribution of electronic states. Our numerical results reveal that rotation removes the degeneracies typically observed in quantum dots while inducing accidental degeneracies at specific angular velocities. Magnetization is observed even in the absence of an external magnetic field, a manifestation of the Barnett effect. Furthermore, the magnetization exhibits oscillatory behavior.
These oscillations resemble the de Haas-van Alphen and Aharonov-Bohm effects observed in systems subjected to magnetic fields. The quantum dot's entropy and heat capacity display distinct dependencies on angular velocity, particularly at high temperatures, where rotational effects become prominent. An adiabatic increase in angular velocity reduces the system's temperature, mimicking the magnetocaloric effect observed in magnetic systems.

Our results extend the understanding of rotational dynamics in confined quantum systems and open new avenues for exploring analogous effects in mesoscopic and nanoscale systems. Future investigations could incorporate additional interactions, such as spin-orbit coupling or anisotropic confinement potentials, to further elucidate the interplay between rotation and other quantum phenomena.

\section*{Acknowledgments}

This work was partially supported by the Brazilian agencies CAPES, CNPq, and FAPEMA. Edilberto O. Silva acknowledges the support from grants CNPq/306308/2022-3, FAPEMA/UNIVERSAL-06395/22, and FAPEMA/APP-12256/22. This study was partly financed by the Coordenação de Aperfeiçoamento de Pessoal de N\'{\i}vel Superior - Brazil (CAPES) - Code 001.

\bibliographystyle{apsrev4-2}

\end{document}